\documentclass[10pt,conference]{IEEEtran}
\IEEEoverridecommandlockouts
\usepackage{cite}
\usepackage{amsmath,amssymb,amsfonts}
\usepackage{algorithm}
\usepackage{algorithmic}
\usepackage{graphicx}
\usepackage{textcomp}
\usepackage{xcolor}
\usepackage{color}
\usepackage{subfigure}

\usepackage{multirow}
\usepackage{caption}
\usepackage{booktabs}
\usepackage{hyperref}
\hypersetup{
    colorlinks=true,
    linkcolor=blue,
    filecolor=magenta,      
    urlcolor=cyan,
    citecolor=blue,
}

\ifCLASSOPTIONcompsoc
  \usepackage[nocompress]{cite}
\else
  \usepackage{cite}
\fi
\ifCLASSINFOpdf
\else
\fi
\begin{document}
%
\title{Joint Task Scheduling and Container Image Caching in Edge Computing}
%
%
%
%

\DeclareRobustCommand*{\IEEEauthorrefmark}[1]{%
    \raisebox{0pt}[0pt][0pt]{\textsuperscript{\footnotesize\ensuremath{#1}}}}

\author{
\IEEEauthorblockN{
Fangyi Mou\IEEEauthorrefmark{1},
Zhiqing Tang\IEEEauthorrefmark{1},
Jiong Lou\IEEEauthorrefmark{2},
Jianxiong Guo\IEEEauthorrefmark{1,3},
Wenhua Wang\IEEEauthorrefmark{1}, and
Tian Wang\IEEEauthorrefmark{1,3}}
\IEEEauthorblockA{\IEEEauthorrefmark{1}Institute of Artificial Intelligence and Future Networks, Beijing Normal University, China}
\IEEEauthorblockA{\IEEEauthorrefmark{2}Department of Computer Science and Engineering, Shanghai Jiao Tong University, China}
\IEEEauthorblockA{\IEEEauthorrefmark{3}Guangdong Key Lab of AI and Multi-Modal Data Processing, BNU-HKBU United International College, China}
\IEEEauthorblockA{fayemoumail@gmail.com, lj1994@sjtu.edu.cn, \{zhiqingtang, jianxiongguo, cs\_wwhua, tianwang\}@bnu.edu.cn}
\thanks{\textit{(Corresponding author: Zhiqing Tang.)}}
}

\maketitle

\begin{abstract}
In Edge Computing (EC), containers have been increasingly used to deploy applications to provide mobile users services. Each container must run based on a container image file that exists locally. However, it has been conspicuously neglected by existing work that effective task scheduling combined with dynamic container image caching is a promising way to reduce the container image download time with the limited bandwidth resource of edge nodes. To fill in such gaps, in this paper, we propose novel joint Task Scheduling and Image Caching (TSIC) algorithms, specifically: 1) We consider the joint task scheduling and image caching problem and formulate it as a Markov Decision Process (MDP), taking the communication delay, waiting delay, and computation delay into consideration; 2) To solve the MDP problem, a TSIC algorithm based on deep reinforcement learning is proposed with the customized state and action spaces and combined with an adaptive caching update algorithm. 3) A real container system is implemented to validate our algorithms. The experiments show that our strategy outperforms the existing baseline approaches by 23\% and 35\% on average in terms of total delay and waiting delay, respectively.
\end{abstract}

\begin{IEEEkeywords}
Task scheduling, image caching, container, edge computing.
\end{IEEEkeywords}

\IEEEdisplaynontitleabstractindextext

%
\IEEEpeerreviewmaketitle

\section{Introduction}\label{sec:introduction}

Edge Computing (EC) has played essential roles in reducing application delays, e.g., real-time face recognition and video surveillance \cite{li2019edge}. Users request different services deployed in nearby edge nodes or the remote cloud. Due to the heterogeneity of the nodes, tasks need to be scheduled effectively to obtain lower communication and computation delay \cite{cai2023joint}. Due to the lightweight and easy-to-deploy features, containers have been widely used to deploy services in EC with container cluster management platforms like KubeEdge or K3s \cite{xiong2018extend}.

Each container must run based on a specific image, containing all the necessary libraries and environments to run an application to provide a specific service. In cloud computing, if the requested image is not stored locally, it can be pulled (i.e., downloaded) through high-speed bandwidth. However, in EC, it is hard to guarantee communication quality due to limited bandwidth, which may lead to a long time to pull the image. It is unacceptable if the image is pulled each time when requested in EC. Besides, due to the limitation of the storage resources of each node, it is impossible to store all the images locally. Therefore, an effective caching algorithm for images can significantly reduce the delay of user tasks in EC. Moreover, the distribution of images also needs to be fully considered when scheduling tasks so that many unnecessary downloads can be avoided.

To perform the task scheduling and image caching efficiently, the following challenges must be solved. First, how to fully extract the complicated edge environment information and make joint decisions. Current researches only consider the limits of the CPU and memory resources of the nodes without the storage resources \cite{zhou2022two}. He \textit{et al.} \cite{he2018s} considered homogeneous services. To fill in such gaps, different types and numbers of images are considered for each node. Existing researchers adopt Reinforcement Learning (RL) algorithms to make joint decisions \cite{qiao2022adaptive, liu2022deep}. However, most of these algorithms are two-time-scale, which make decisions separately, and environmental information cannot be fully considered. In real systems, the state space is large and sparse due to the heterogeneity of tasks and the complexity of system states. Therefore, a state-sharing and multi-action RL algorithm is proposed to use limited system state information better. A Q-network composed of multiple parts is designed, and both scheduling actions (decisions) and caching actions are output by this Q-network while training with different rewards. 

Then the second challenge is how to fully consider the storage limit and image distribution when making decisions. Some existing research on service placement in EC has considered the storage resource limitation of each node \cite{chu2023joint, wang2022joint}, but when and which image needs to be removed are not considered when the storage capacity is insufficient. Generally, image cache removal is based on the Least Frequently Used (LFU) algorithm, in which the frequencies of different types of images are maintained with a fixed-size LFU memory \cite{drolia2017cachier}. However, since the file size of each image is different, it is not reasonable to consider the frequency with a fixed size in EC. To solve these problems, an adaptive LFU-based caching update algorithm is proposed, which is storage-aware, variable-size, and container image size-weighted.

In this paper, we first model the joint task scheduling and image caching problem as a Markov Decision Process (MDP) to minimize the communication and computation delay of the tasks. Based on this, a Deep Q-Learning (DQL) based algorithm is proposed with a shared state and multiple action spaces combined with the adaptive LFU algorithm \cite{drolia2017cachier}. Finally, we implement our algorithms in a real container system consisting of computers and Raspberry Pis. The deployment of containers and images is controlled by the Docker Python client and socket server \cite{dockerpython, socketserver}, and a controller node is deployed to make task scheduling and image caching decisions. Experimental results show that our algorithms perform better than the baselines.

To sum up, the contributions of this paper are as follows:
\begin{itemize}
    \item We first consider the joint task scheduling and image caching problem and formulate it as an MDP, which aims to minimize communication, waiting, and computation delays. The heterogeneity of nodes and services is taken into consideration.
    \item To solve this problem, a joint task scheduling and image caching algorithm based on DQL is proposed, which includes a state-sharing multi-action Q-network and combined with an adaptive LFU caching update algorithm.
    \item Our algorithms are implemented in a real container system. The experimental results show that our strategy outperforms the baseline approaches by 23\% and 35\% on average in terms of total delay and waiting delay, respectively.
\end{itemize}

\section{Related Work}
\label{section_related_work}

\subsection{Joint Optimization}

There have been many researches on joint task scheduling and service placement optimization problems in EC \cite{cai2023joint, zhou2022two, he2018s, chu2023joint, wang2022joint}. Some algorithms have been proposed to reduce the overall latency effectively by jointly task offloading and service caching. Li \textit{et al.} \cite{li2023esmo} propose a joint frame scheduling and model caching algorithm and deploy a target recognition prototype to evaluate the performance. Zhang \textit{et al.} \cite{zhao2023silod} propose a scheduling framework with an enhanced job performance estimator that co-designs the cluster scheduler and the cache subsystems for deep learning training. Xiao \textit{et al.} \cite{xiao2022multi} aim to jointly optimize parallel task offloading and content caching to minimize task delay and energy consumption. Kamran \textit{et al.} \cite{kamran2021deco} present a framework for jointly optimizing computation scheduling, caching, and request forwarding in EC to enhance average task completion time performance. Fan \textit{et al.} \cite{fan2022joint} propose a resource management scheme that jointly optimizes task offloading and service caching to maximize energy consumption benefits. However, they ignore the chance that total delays can be further optimized by jointly considering task scheduling and image caching in an online manner.

\subsection{Reinforcement Learning}

Some algorithms have been proposed based on RL to solve the task scheduling problem in EC. Tang \textit{et al.} \cite{tang2023layer} propose an RL-based layer-aware task scheduling algorithm to minimize the task completion time. Chou \textit{et al.} \cite{chou2021pricing} address the user association and video quality selection problem and propose a deep deterministic policy gradient-based algorithm. Tang \textit{et al.} \cite{tang2022collective} introduce a collective deep RL algorithm to optimize intelligence sharing policies. Gao \textit{et al.} \cite{gao2022large} propose a decentralized computation offloading solution based on the attention-weighted recurrent multi-agent actor-critic for latency-sensitive tasks. Besides, there have been some researches using RL for joint decision-making. Al-Abiad \textit{et al.} \cite{al2022joint} combine RL with cross-layer network coding to optimize caching strategy and coding decisions. Qiao \textit{et al.} \cite{qiao2022adaptive} employ deep RL for client selection and local iteration number decisions to enhance content caching. Liu \textit{et al.} \cite{liu2022deep} propose an approach that employs parameterized deep Q networks to make joint decisions on service placement and computation resource allocation to minimize the total latency of tasks.

\section{System Model and Problem Formulation}
\label{section_system_model}


\subsection{System Model}
\label{subsection_system_model}

We consider an EC system that includes a set of edge nodes $\mathbf{N} = \{n_1, n_2, ..., n_{|\mathbf{N}|}\}$, a set of tasks $\mathbf{U} = \{u_1, u_2, ..., u_{|\mathbf{U}|}\}$, a set of services $\mathbf{V} = \{v_1, v_2, ... v_{|\mathbf{V}|}\}$, and a set of images $\mathbf{M} = \{m_1, m_2, ..., m_{|\mathbf{M}|}\}$. The nodes provide different kinds of services. Tasks are generated from users and sent to the nodes to be processed. The images are located in a remote cloud or nearby nodes. 

A node $n \in \mathbf{N}$ is located at location $l_n$. Each node has its CPU $p_n$, memory $e_n$, and storage $z_n$. A list of available images $\mathbf{M}_n(t)$ is also maintained on each node. Generally, the available resources at time slot $t$ is denoted as $p_n(t)$, $e_n(t)$, and $z_n(t)$, respectively. There are $|\mathbf{M}|$ types of different images, and at most one copy of each image is needed on each node since multiple service instances providing the same service are run based on the same image. As a result, the image list $\mathbf{M}_n(t)$ can be denoted as $\mathbf{M}_n(t) = [x_n^1, x_n^2, ..., x_n^{|\mathbf{M}|}]$, where $x_n^m \in \{0, 1\}$. If $x_n^m = 1$, the image $m \in \mathbf{M}$ exists on node $n$. Otherwise, it does not exist on this node.

Furthermore, each task $u \in \mathbf{U}$ has its requested service $v_u \in \mathbf{V}$, data size $z_u$, and the location $l_u$. The requested service is denoted as $v_u = [y_u^1, y_u^2, ..., y_u^{|\mathbf{V}|}]$, where $y_u^v \in \{0, 1\}$ denotes the request for each service type. If $y_u^v = 1$, task $u$ is requesting service $v$. It is assumed that each task only requests one service. Thus $\sum_{v \in\mathbf{V}}y_u^v = 1$. Some specific data needs to be transmitted to the node to be processed for each task $u$, and $z_u$ is used to denote the size of this data. The location of the task is denoted as $l_u$. For each node $n$, the set of users it serves is denoted as $\mathbf{U}_n$.

To process the user requests, different services are provided on each node. Each service $v \in \mathbf{V}$ runs based on an image $m$, and multiple containers providing the same service can run simultaneously based on the same image to increase the service capacity. When starting a service $v$, the corresponding image $m$ must exist locally on the node. If not, the image should be pulled first. Each image $m$ has its file size $z_m$.

\subsection{Cost and Constraints}
\label{subsection_cost_model}

In EC, the delay for task execution is significant to the users' Quality of Service (QoS) \cite{tang2018migration}. So the system cost is defined as the total delay $d_{un}$ of task executions, which is defined as:
\begin{equation}
    d_{un} = d_{un}^{comm} + d_{un}^{wait} + d_{un}^{comp},
\end{equation}
where $d_{un}^{comm}$ denotes the communication delay. $d_{un}^{wait}$ is the waiting time for service initialization, which includes the pulling time of the image and the starting time of the service. $d_{un}^{comp}$ is the computation time for task processing. 


There are also some constraints during scheduling. The communication constraint of node $n$ means that the total bandwidth usage should not exceed the bandwidth of node $n$, which is denoted as:
\begin{equation}
\label{eq:constraint1}
    \sum_{u \in \mathbf{U}_n} b_u \leq b_n,
\end{equation}
where $b_u$ is the bandwidth consumption of $u$, and $b_n$ is the bandwidth capacity of node $n$. Besides, for node $n$, the CPU and memory resources are limited as:
\begin{equation}
\label{eq:constraint2}
\begin{aligned}
    \sum_{u \in \mathbf{U}_n} p_u \leq p_n, 
    \sum_{u \in \mathbf{U}_n} e_u \leq e_n,
\end{aligned}
\end{equation}
where $p_u$ and $e_u$ is the CPU and memory consumption of task $u$. One more important constraint that current work should have seriously considered is the storage constraint of each node. In EC, the storage capacity of each node is limited, e.g., the general storage capacity of a Raspberry Pi is usually 8GB or 16GB, which is not large enough to store all necessary images locally (the container size is usually a few hundred MB). The storage constraint is then defined as:
\begin{equation}
\label{eq:constraint3}
    \sum_{u \in \mathbf{U}_n} z_u + \sum_{m \in \mathbf{M}} x_n^m \times z_m \leq z_n,
\end{equation}
where $z_u$ is the size of data received from task $u$.

\subsection{Problem Formulation}
\label{subsection_problem_formulation}

In this paper, we aim to minimize the total delay of tasks with the constraints. The problem is then defined as follows:

\newtheorem{problem}{Problem}
\begin{problem}
\label{pro:1}
\begin{equation}
\begin{aligned}
\label{eq:constraint}
 &\min \sum_{u \in \mathbf{U}} \sum_{n \in \mathbf{N}} (d_{un}^{comm} + d_{un}^{wait} + d_{un}^{comp}) \\
& s.t. \quad \text{Eqs}. \ (\ref{eq:constraint1}), (\ref{eq:constraint2}), (\ref{eq:constraint3}),
\end{aligned}
\end{equation}
\end{problem}

Problem \ref{pro:1} is an advanced bin-packing problem, which is NP-hard and can only be solved heuristically. However, most existing heuristic algorithms cannot solve the large-scale problem in a real-world environment. In this problem, $d_{un}^{comm}$, $d_{un}^{wait}$, and $d_{un}^{comp}$ can be represented as the addition of $d_{un}^{comm}(t)$, $d_{un}^{wait}(t)$, and $d_{un}^{comp}(t)$ during each time slot $t$. Besides, the first-order transition probability of the users' resource demands is also quasi-static for a long period and non-uniformly distribution by properly choosing the time slot duration \cite{tang2018migration}, which is a sequential decision-making process and has memoryless property. Therefore, this problem can be modeled as an MDP. To solve this problem, RL-based algorithms are adopted.


\section{Algorithms}
\label{section_algorithms}


\subsection{Reinforcement Learning Settings}
\label{subsection_reinforcement}

In RL algorithms, at each time $t$, the RL agent collects system state $s_t$ and calculates the reward during the last time slot $r_{t}$. Then, the agent selects an action $a_t$ according to a pre-defined strategy. After performing the action, the system would transit to the new state $s_{t+1}$ in the next time slot. Similarly, the RL agent will repeat the above operations, i.e., calculating reward $r_{t+1}$ and selecting new action $a_{t+1}$ according to $s_{t+1}$. 

Among all kinds of RL algorithms, Q-learning \cite{watkins1992q} has an advantage in fast computation, which is consistent with the requirement of rapid decision-making in EC. The quality of each state-action pair is indicated by Q-value $Q(s_t, a_t)$. The RL agent tends to select an action with a larger Q-value each time. The $Q(s_{t}, a_{t})$ can be updated with the learning rule:
\begin{equation}
    \begin{aligned}
    \label{eq:qlearning}
    Q(s_{t}, a_{t}) \leftarrow& (1 - \alpha) Q(s_{t}, a_{t}) \\ 
    &+ \alpha \left[ r_{t} + \gamma \times \max_{a_{t+1}}Q(s_{t+1}, a_{t+1})\right],
    \end{aligned}
\end{equation}
where $\alpha$ is the learning rate and $\gamma$ is the discount parameter.

\textbf{State}: The task scheduling decisions are made based on the available resource capacity of all nodes and the features of the coming task. The state of available resource for node $n$ at time $t$ can be denoted as $s_t^{n} =  [p_n(t), e_n(t), z_n(t), \mathbf{M}_n(t)]$. Then, the state of all nodes can be denoted as $s_t^{\textbf{N}} = \{s_t^{n} | n \in \textbf{N}\}$. Besides, for task $u$, the state can be denoted as $s_t^{u} = [v_u, z_u, l_u]$. Then the state of the task scheduling decision can be defined as $s_t^{s} = \left[s_t^{\textbf{N}}, s_t^{u}\right] \in \mathbf{S^s}$, where $\mathbf{S^s}$ is the set of all scheduling states.

Furthermore, the distribution of all existing requests is also needed for image caching decisions to determine the popularity of images better and make more appropriate decisions. The state of request distribution $s_t^r$ can be denoted as:
\begin{align}
s_t^r = 
\begin{bmatrix}
w_{1,1}(t)      & \cdots & w_{1,|\mathbf{M}|}(t)      \\
\vdots & \ddots & \vdots \\
w_{|\mathbf{N}|,1}(t)      & \cdots & w_{|\mathbf{N}|,|\mathbf{M}|}(t)
\end{bmatrix}, 
\end{align}
where $w_{n, v}$ is the number of requests of service $v$ on node $n$. Then, the state of the image caching decision can be denoted as:
\begin{equation}
\label{eq:s_tc}
    s_t^{c} = [s_t^{\mathbf{N}}, s_t^{u}, s_t^r] = [s_t^s, s_t^r] \in \mathbf{S^c},
\end{equation}
where $\mathbf{S^c}$ is the set of all caching states. The state of task scheduling $s_t^s$ is shared with the state of image caching. In this way, the utilization of information and decision-making accuracy can be effectively improved.


\textbf{Action}: The decision of task scheduling is made to select a node for task processing, which can be denoted as $a_t^s \in \mathbf{A^s} =  \mathbf{N}$. Moreover, the best action can be obtained as follows:
\begin{equation}
\label{eq:action_selection_schedule}
    a_t^{s*} = \arg\max_{a_t^s} \left(Q(s_t^s, a_t^s) | (\mathbf{M}_{a_t^s}(t) * v_{u_t} = 1)\right),
\end{equation}
where $\mathbf{M}_{a_t^s}(t) * v_{u_t}$ is defined as follows:
\begin{equation}
    \mathbf{M}_{a_t^s}(t) * v_{u_t} = x_t^1 y_t^1 + x_t^2 y_t^2 ... + x_t^{|\mathbf{M}|} y_t^{|\mathbf{M}|}.
\end{equation}
If $\mathbf{M}_{a_t^s}(t) * v_{u_t} = 1$, then the image of requested service exists locally on node $a_t^s$. In short, the best action is to select a suitable node with the requested image stored locally. Random action is selected if the requested image is not located on any node.


Besides, when updating the image caching, the goal is to select the most popular image to be cached on a node. Then the action of image caching is defined as $a_t^c = c_{m, n} \in \mathbf{A^c}$, where $c_{m, n}$ means that the image $m$ is selected to be cached on node $n$. The action of time slot $t$ is finally defined as $a_t = [a_t^s, a_t^c]$. Scheduling and caching actions are selected based on $\epsilon$-greedy algorithm \cite{tang2018migration}.

\textbf{Reward}: Task scheduling and image caching rewards are also defined separately. For task scheduling, we aim to minimize the total delay, so the reward is set to the negative of the total delay, which is defined as:
\begin{equation}
\label{eq:reward1}
r_t^s = - d_{un}.
\end{equation}
In our real container system, the reward is collected asynchronously after the task is completed.

Furthermore, image caching aims to find the most popular image and deploy it to an appropriate node. The request number of the image can represent the popularity of the image after a caching decision is made. Therefore, the actions that have a larger Q-value but have not been selected (unscheduled actions) are recorded, which is denoted as:
\begin{equation}
    \mathbf{A_t^u} = \{a_t^u | Q(s_t^s, a_t^u) > Q(s_t^s, a_t^s)\}.
\end{equation}
If there is no unscheduled action, then $\mathbf{A_t^u} = \emptyset$. A popularity matrix $g_t$ for the task $u$ at time $t$ can be obtained, which is defined as:
\begin{align}
g_t = 
\begin{bmatrix}
g_{1,1}(t)      & \cdots & g_{1,|\mathbf{M}|}(t)      \\
\vdots & \ddots & \vdots \\
g_{|\mathbf{N}|,1}(t)      & \cdots & g_{|\mathbf{N}|,|\mathbf{M}|}(t)
\end{bmatrix}, 
\end{align}
where
\begin{equation}
    g_{m,n}(t) =
    \begin{cases}
    1, & n \in \mathbf{A_t^u} \cap a_t^s \ \text{and}\ l_u = v \\
    0, & \text{Otherwise}.
    \end{cases}
\end{equation}

The popularity matrix $g_t$ is stored in a caching memory $D_c$ along with the state and caching action for future reward calculation and network training, i.e., $(s_t, c_{m, n}, g_t)$ is stored to $D_c$ at each time slot $t$. The reward is then calculated several time slots after the caching action is made, e.g., every $T_c$ time slot. Then the reward can be defined as:
\begin{equation}
\label{eq:reward2}
    r_t^c = \sum_{t}^{t + T_c} g_{m, n}(t).
\end{equation}
Similarly, the reward in time $t$ is denoted as $r_t = [r_t^s, r_t^c]$.

\subsection{State-Sharing Multi-Action Scheduling Algorithm}
\label{subsection_drl_based}

\begin{figure}[!t]
    \centering
    \includegraphics[width=1\linewidth]{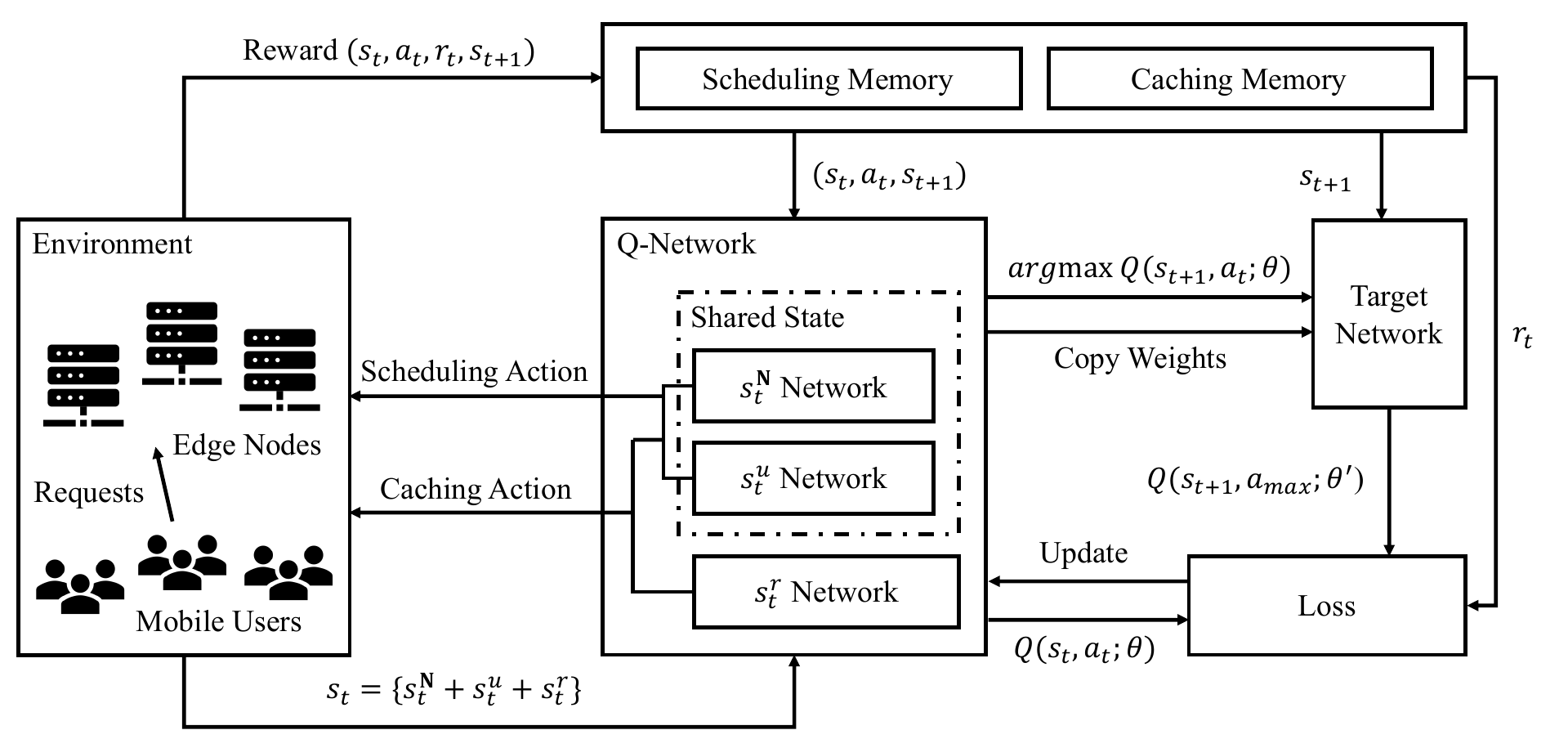}
    \caption{State-Sharing Multi-Action Scheduling Algorithm}
    \label{fig:dqn}
\end{figure}

\begin{algorithm}
\caption{Q-Network Update}
\label{alg:qnetwork}
\begin{algorithmic}[1]
\renewcommand{\algorithmicrequire}{\textbf{Input:}}
\renewcommand{\algorithmicensure}{\textbf{Output:}}
\REQUIRE
$D_s, D_c$
\ENSURE
$\theta$
\FOR{$D \in \{D_s, D_c\}$}
\STATE Sample $D_t \subset D$
\FOR {$\left( s_{t}^{(j)},a_{t}^{(j)},r_{t}^{(j)},s_{t+1}^{(j)}  \right)$ in $D_t$}
\STATE Calculate $y_{d}(t)$ by Eq. (\ref{eq:yd})
\STATE Calculate $L_\theta$ by Eq. (\ref{eq:lossfunction})
\ENDFOR
\STATE $\theta = arg \min_{\theta} L_\theta$
\STATE Return $\theta$
\ENDFOR
\end{algorithmic}
\end{algorithm}

The process of the state-sharing multi-action scheduling algorithm is illustrated in Fig. \ref{fig:dqn}. During each decision, firstly, the system state $s_t$ is obtained from EC environment, which includes the $s_t^{\mathbf{N}}$, $s_t^{u}$ and $s_t^{r}$. Secondly, the scheduling and caching actions are obtained from the Q-network with the state. In the Q-network, the sub-networks, which include the $s_t^{\mathbf{N}}$-network and $s_t^{u}$-network, are shared for scheduling and caching decisions since this information is needed for both decisions. Besides, the training data can be used more efficiently by sharing these sub-networks. After that, the environment returns the reward $r_t$ and the tuple $(s_t, a_t, r_t, s_{t+1})$ is stored in the replay memory, where $a_t = [a_t^s, a_t^c]$, and $r_t = [r_t^s, r_t^c]$. The tuples $(s_t, a_t^s, r_t^s, s_{t+1})$ and $(s_t, a_t^c, r_t^c, s_{t+1})$ are stored in scheduling memory $D_s$ and caching memory $D_c$, respectively, for further Q-network update. Finally, the loss is calculated, and the Q-network is updated with Algorithm \ref{alg:qnetwork}.

In Algorithm \ref{alg:qnetwork}, for each memory $D \in \{D_s, D_c\}$, first a subset $D_t$ is sampled from $D$ and used to train the Q-network. For each entry $\left( s_{t}^{(j)},a_{t}^{(j)},r_{t}^{(j)},s_{t+1}^{(j)}  \right)$ in $D_t$, the corresponding $y_{t}$ is calculated according to Eq. (\ref{eq:yd}). Then the loss $L_\theta$ is calculated by Eq. (\ref{eq:lossfunction}) and used to update the weights. Finally, a gradient descent step is performed in the training network, and the weights $\theta$ of the current network are occasionally copied to the target network. 

The historical information is stored in replay memory $[D_s, D_c]$. The Q-value is denoted as $Q(s_t, a_t; \theta)$ with network weight $\theta$. The objective of the training is to minimize the loss function $L_\theta$, which is defined as:
\begin{equation}
\label{eq:lossfunction}
L_\theta = (y_t - Q(s_{t}, a_{t};\theta))^2.
\end{equation}


However, only one DNN works as a whole leads to a problem that the target is likely to shift with each update. To overcome this problem, a target network \cite{mnih2015human} is used, which provides stable $Q(s_{t}, a_{t};\theta')$. Moreover, $y_t$ is defined as:
\begin{align}
\label{eq:y}
 y_t &= r_{t} + \gamma \max Q(s_{t+1}, a_{t+1};\theta'),
\end{align}
where $\theta'$ is the weight of the target network, which is reset to $\theta$ from time to time. Furthermore, the max operator $\max Q(s_{t+1}, a_{t+1};\theta')$ in Eq. (\ref{eq:y}) uses the same Q-values both to choose and to evaluate an action. This makes it more likely to select overestimated values. To solve this problem, two Q-value functions are learned by assigning experiences randomly to update one of them in double Q-learning \cite{van2016deep}. In Algorithm \ref{alg:qnetwork}, the policy network can be used to evaluate the greedy policy, and the target network can be used to estimate its value \cite{van2016deep}. The target $y_t$ is then revised as follows:
\begin{align}
\label{eq:yd}
y_t &= r_{t} + \gamma Q(s_{t+1}, arg\max_{a} Q(s_{t+1},a, \theta);\theta').
\end{align}

With Algorithm \ref{alg:qnetwork}, the scheduling decision and caching decision (for pulling the image) can be made. Then the adaptive LFU-based caching algorithm (for removing the image) is described as follows.

\begin{algorithm}
\caption{Adaptive LFU-based Caching Update}
\label{alg:lfu}
\begin{algorithmic}[1]
\renewcommand{\algorithmicrequire}{\textbf{Input:}}
\renewcommand{\algorithmicensure}{\textbf{Output:}}
\REQUIRE
$E, a_t^c = c_{m,n}, D_l(t-1)$
\ENSURE
$D_l(t)$
\STATE Set $\mathbf{M}_t = \emptyset$
\IF{$m \notin \mathbf{M}_n(t)$}
\WHILE{$z_m + \sum_{m' \in \mathbf{M}_n(t)} z_{m'} > z_n$}
\STATE Add image $m_t$ to $\mathbf{M}_t$ by Eq. (\ref{eq:it})
\STATE Remove image $m$ from $\mathbf{M}_n(t)$
\ENDWHILE
\STATE Remove all images in $\mathbf{M}_t$
\STATE Pull image $m$
\ENDIF
\STATE Update priority of image $m$ by Eq. (\ref{eq:iip})
\STATE Update LFU memory $D_l(t)$
\STATE Return $D_l(t)$
\end{algorithmic}
\end{algorithm}

\subsection{Adaptive LFU-based Caching Update Algorithm}
\label{subsection_lru_based}

The adaptive LFU-based caching update algorithm is shown in Algorithm \ref{alg:lfu}. The input is the EC environment $E$, caching decision $a_t^c = c_{m, n}$, and the LFU memory $D_l(t-1)$. The output is the updated $D_l(t)$. The LFU memory $D_l(t)$ contains the memory $D_l^n(t)$ for each node $n$, i.e., $D_l(t) = \{D_l^n(t)| n \in \mathbf{N}\}$. Each $D_l^n(t)$ is an ordered dictionary that records the frequency of each existing image $m \in \mathbf{M}_n(t)$.


As shown in line 1 of Algorithm \ref{alg:lfu}, first, an image set $\mathbf{M}_t$ is initialized, which is used to record the images that need to be removed. Then, as shown in lines 2 - 9, if the request image $m$ is not located in node $n$, it needs to be pulled. Instead of a fixed-size LFU memory, the LFU memory is maintained according to the available storage capacity of each node. If there is not enough available capacity, an image $I_t$ is removed according to the priority $h_{mn}$, which is defined as:
\begin{equation}
\label{eq:iip}
    h_{mn} = f_{mn} \times z_m,
\end{equation}
where $f_{mn}$ is the frequency of image $m$ on node $n$. The priority $h_{mn}$ is related to the size of the image. This is reasonable since pulling a larger image takes more time, making us tend not to remove large files frequently. The image with minimal priority is removed, which is denoted as:
\begin{equation}
\label{eq:it}
    m_t = \arg \min_{m} (h_{mn} | m \in \mathbf{M}_n(t)).
\end{equation}

After removing all necessary images, the requested image $m$ is then pulled. Finally, as shown in lines 10 - 11, the priority of image $m$ is updated, and the LFU caching memory is updated.

\begin{algorithm}
\caption{TSIC}
\label{alg:joint}
\begin{algorithmic}[1]
\renewcommand{\algorithmicrequire}{\textbf{Input:}}
\renewcommand{\algorithmicensure}{\textbf{Output:}}
\REQUIRE
$u$
\ENSURE
$a_t^s, a_t^c$
\FOR{$t \in [1, T]$}
\IF{$msg = \text{request}$}
\STATE Get state $s_t$, action $a_t^s$, and action $a_t^c$
\STATE Push $(s_t, a_t^s)$ to $D_s^{tmp}$, push $(s_t, a_t^c)$ to $D_c^{tmp}$
\STATE Start the requested service on $a_t^s$
\STATE Send $a_t^s$ to $u$
\IF{$t \ \%$ CachingUpdate $= 0$ }
\STATE Call Algorithm \ref{alg:lfu} to update caching
\ENDIF
\ELSIF{$msg = \text{reward}$}
\STATE Get reward $r_t = [r_t^s, r_t^c]$ by Eqs. (\ref{eq:reward1}), (\ref{eq:reward2})
\STATE Get current state $s_{t+1}$
\STATE Get $(s_t,a_t^s)$ from $D_s^{tmp}$, $(s_t,a_t^c)$ from $D_c^{tmp}$
\STATE Push $(s_t, a_t^s, r_t^s, s_{t+1})$ to $D_s$
\STATE Push $(s_t, a_t^c, r_t^c, s_{t+1})$ to $D_c$
\STATE Call Algorithm \ref{alg:qnetwork} to update the Q-network
\IF{$t \ \%$ TargetNetworkUpdate $= 0$}
\STATE Set $\theta' = \theta$
\ENDIF
\ENDIF
\ENDFOR
\end{algorithmic}
\end{algorithm}

\subsection{Joint Task Scheduling and Image Caching Algorithm}
\label{subsection_joint}

The joint Task Scheduling and Image Caching (TSIC) algorithm is shown in Algorithm \ref{alg:joint}. The input is the task $u$ with a message $msg$ and some other features defined in Subsection \ref{subsection_system_model}. The message $msg$ is used to denote the request type since the reward cannot be obtained in time, and it is collected asynchronously. The output is the scheduling decision $a_t^s$ and caching decision $a_t^c$. This algorithm runs on the controller node with relatively sufficient computation resources. An agent on the controller node is responsible for handling the requests.

At each time slot $t \in [1, T]$, the agent receives the request from a new task $u$ and extracts the message $msg$. As shown in lines 2 - 9, if the request type is `request', the agent first gets the state $s_t$, action $a_t^s$, and action $a_t^c$. Then, the agent pushes the states and actions to temporary memories $D_s^{tmp}$ and $D_c^{tmp}$ and starts the requested service on node $a_t^s$. After that, the agent sends the scheduling decision to the task. Furthermore, if the container image caching needs to be updated on the node, Algorithm \ref{alg:lfu} is called to update the caching.

Besides, as shown in lines 10 - 20, if the request type is `reward', the reward is calculated according to Eqs. (\ref{eq:reward1}) and (\ref{eq:reward2}). The reward tuples $(s_t, a_t^s, r_t^s, s_{t+1})$ and $(s_t, a_t^c, r_t^c, s_{t+1})$ are pushed to corresponding replay memory. After that, the Q-network is updated according to Algorithm \ref{alg:qnetwork}. Finally, the target network is updated from time to time.

\section{System Implementation}
\label{subsection_system}

We have implemented a prototype container system to validate the effectiveness of our algorithms.

\subsection{System Workflow} 

\begin{figure}[!t]
    \centering
    \includegraphics[width=0.9\linewidth]{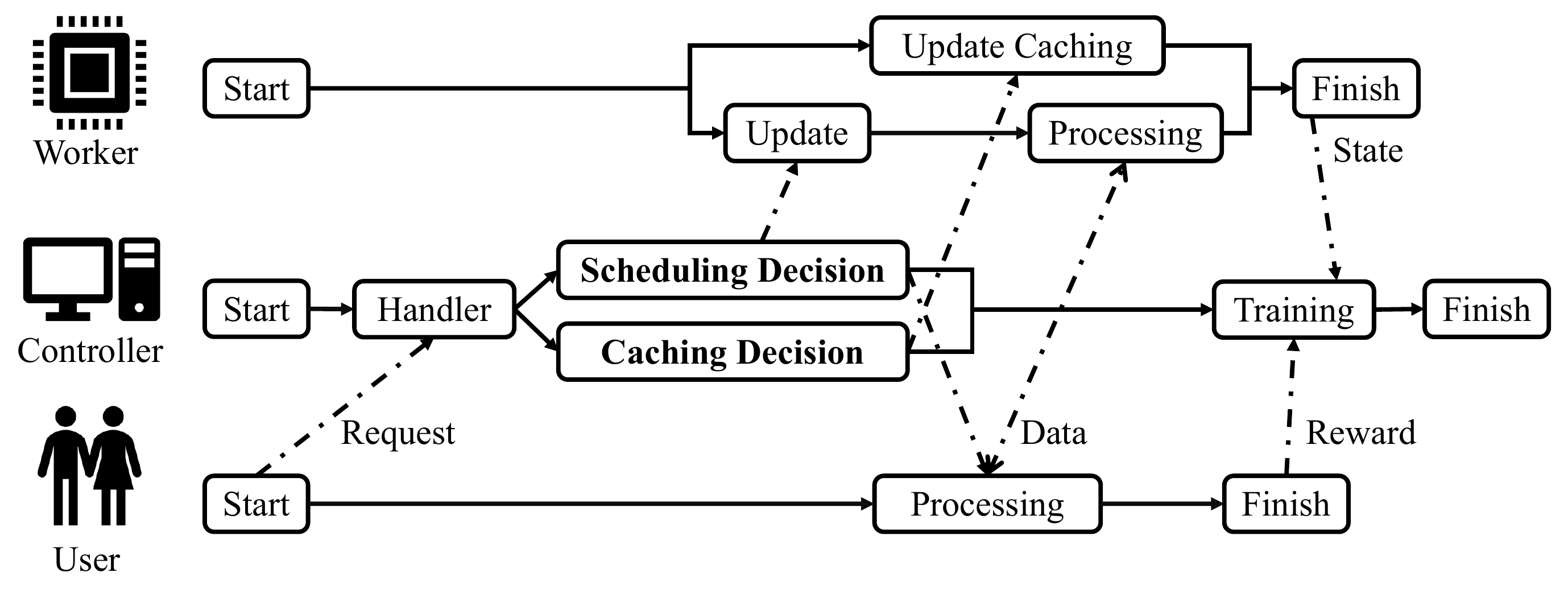}
    \caption{System Workflow}
    \label{fig:process}
\end{figure}

The process of one task request is shown in Fig. \ref{fig:process}. First, all the components are started, and the user sends the request to the controller node. The controller node receives the request and makes the scheduling and caching decisions through the agent. Then, the scheduling decision is sent to the selected worker node to update the service status (e.g., pull necessary images and start the requested service).

Meanwhile, the scheduling decision is also sent back to the user. The user then sends the necessary data to the corresponding worker node to be processed. Furthermore, the caching decision is used to update the caching of the node from time to time. Finally, the agent on the controller node collects the scheduling decision, caching decision, reward, and state to train the Q-network.

\subsection{Main Components}

The main components of the system include a controller node, several worker nodes, and a list of users. The communications among different nodes and users are implemented by Python socket server \cite{socketserver}. The operations of containers and images are implemented with the Docker Python API \cite{dockerpython}.

\textbf{Controller Node}: A PC with an i7-8700 CPU, 16 GB ram, and Ubuntu 18.04 OS is used as the controller node. The main functions are as follows:

\begin{enumerate}
    \item $handle$: Handle user requests according to different request types described in Algorithm \ref{alg:joint}.
    \item $update\_node$: Send image-related command to the corresponding node to update the image caching, including the pull and removal of images.
    \item $get\_scheduling\_decision$: Get the scheduling decisions from the TSIC algorithm or other baselines.
    \item $get\_caching\_decision$: Get the caching decisions.
    \item $network$: The component of Q-network.
    \item $memory$: The replay memory.
\end{enumerate}

\textbf{Worker Node}: The worker nodes are a set of Raspberry Pi 3 Model B+, with Cortex-A53 CPU, 1 GB ram, and 8 GB (or 16 GB, 32 GB) Micro SD Card inserted. Different storage spaces of Raspberry Pis bring the heterogeneity of nodes. The main functions of worker nodes are described as follows:

\begin{enumerate}
    \item $handle$: The core function processes the requests from the controller node or users. The command type includes initializing the node, updating the images, and processing the request.
    \item $init\_node$: Initialize the worker nodes, which includes collecting the present image information and sending the initial state to the controller node.
    \item $get\_state$: Collect the node state, including the available CPU, memory, storage space, and image list.
    \item $send\_state$: Send node state to the controller node.
    \item $update\_image$: Pull or remove the specific images.
    \item $check\_image$: Check if an image exists on this node.
\end{enumerate}

\textbf{User}: Another PC is used to simulate a group of users. The requested service type is generated based on random distribution. The main functions are illustrated as follows:

\begin{enumerate}
    \item $get\_user\_list$: Generate the user list based on random distribution.
    \item $send\_request$: Send each user request to the controller.
    \item $send\_data$: Send the user data to the scheduled worker node to be processed.
    \item $send\_reward$: Send the reward to the controller node.
\end{enumerate}

\section{Performance Evaluation}
\label{section_performance}

This section introduces the experimental settings and results in \ref{subsection_parameter} and \ref{subsection_experimental}, respectively.

\begin{figure}[!t]
    \centering
    \includegraphics[width=0.85\linewidth]{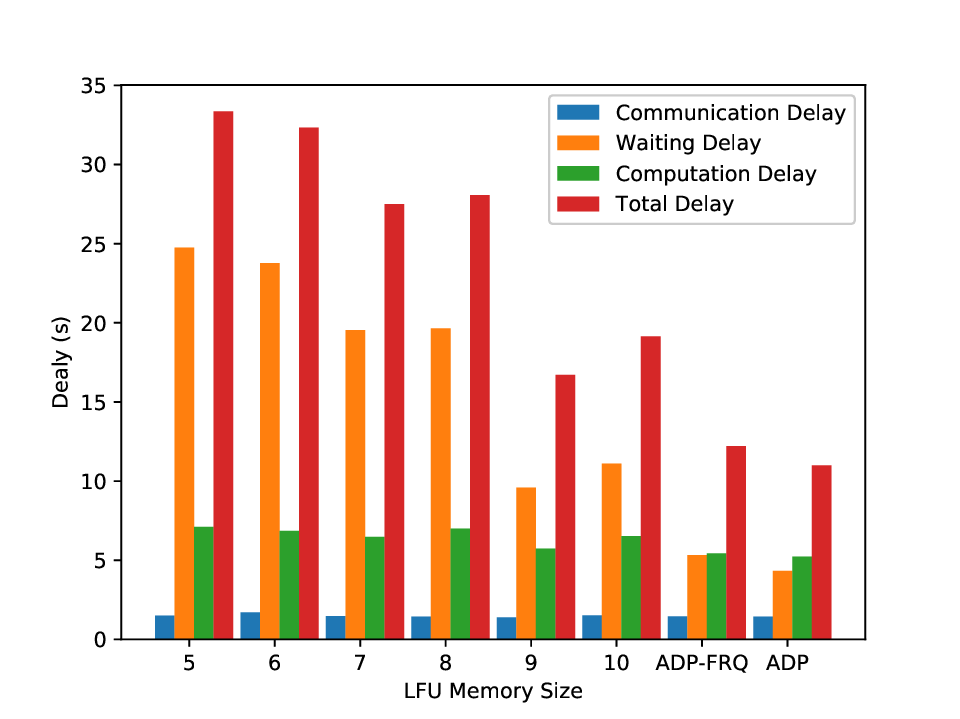}
    \caption{Delay with Different LFU Memory Size}
    \label{fig:lfu_size}
\end{figure}

\begin{figure*}[t]
	\centering
	\subfigure[Communication Delay]{
  \begin{minipage}[t]{0.22\textwidth}
    \centering
    \includegraphics[width=1\textwidth]{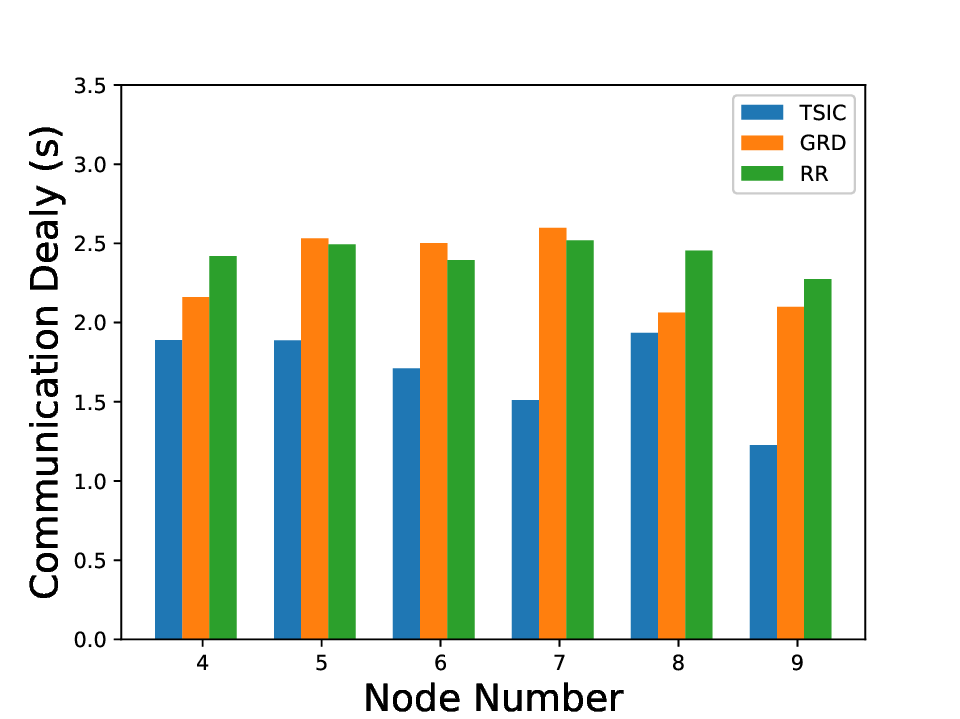}
    \label{fig:node_number_communication_delay}
  \end{minipage}
  }
	\hspace{\fill}
	\subfigure[Waiting Delay]{
	\begin{minipage}[t]{0.22\textwidth}
		\centering
		\includegraphics[width=1\textwidth]{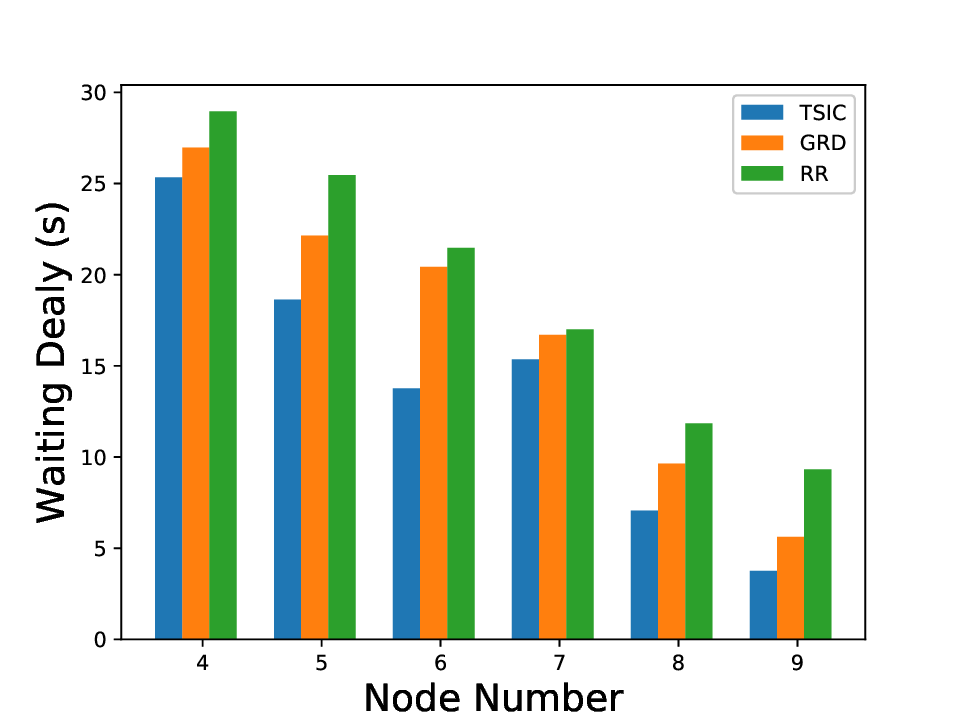}
		\label{fig:node_number_waiting_delay}
	\end{minipage}
        }
	\hspace{\fill}
	\subfigure[Computation Delay]{
  \begin{minipage}[t]{0.22\textwidth}
    \centering
    \includegraphics[width=1\textwidth]{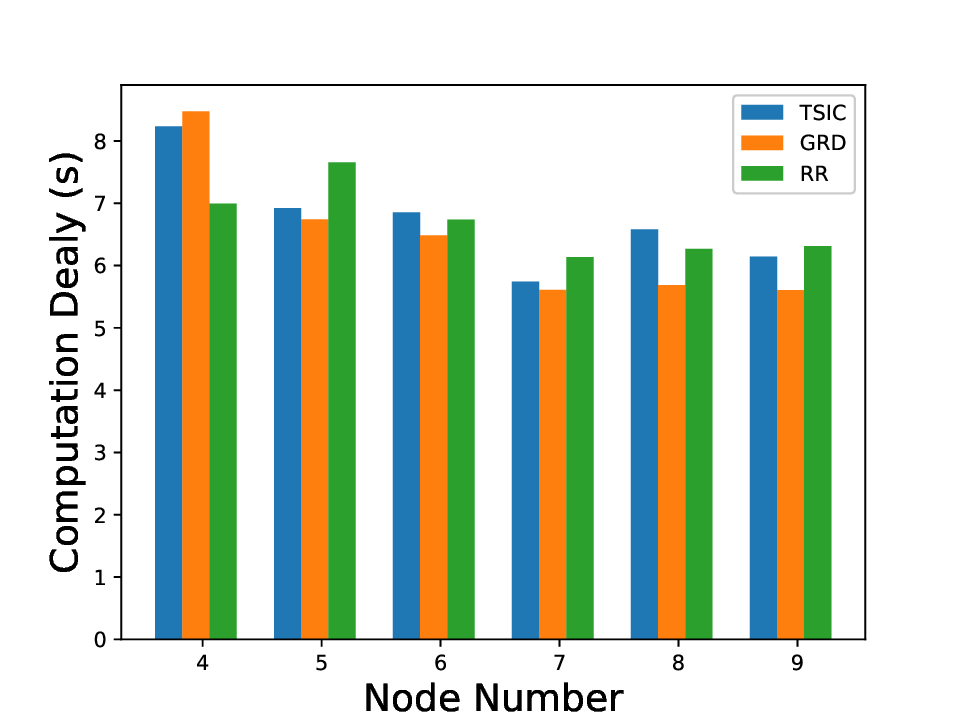}
    \label{fig:node_number_computation_delay}
  \end{minipage}
  }
   \hspace{\fill}
	\subfigure[Total Delay]{
  \begin{minipage}[t]{0.22\textwidth}
    \centering
    \includegraphics[width=1\textwidth]{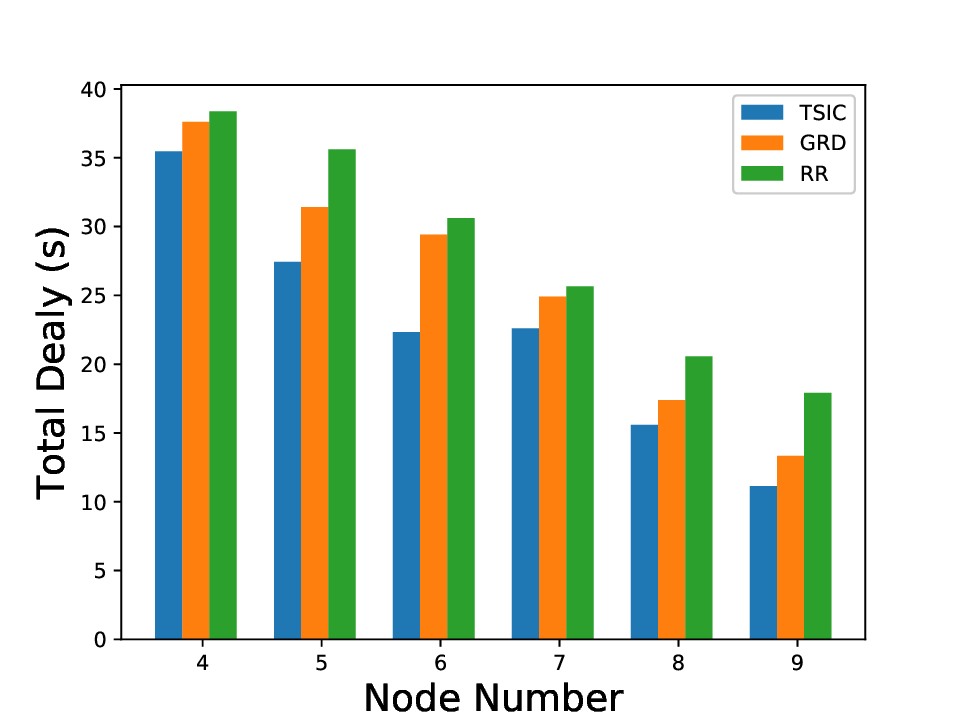}
    \label{fig:node_number_total_delay}
  \end{minipage}
  }
	\caption{Performance with Different Node Number}
	\label{fig:node_number}
\end{figure*}

\begin{figure*}[t]
	\centering
	\subfigure[Communication Delay]{
  \begin{minipage}[t]{0.22\textwidth}
    \centering
    \includegraphics[width=1\textwidth]{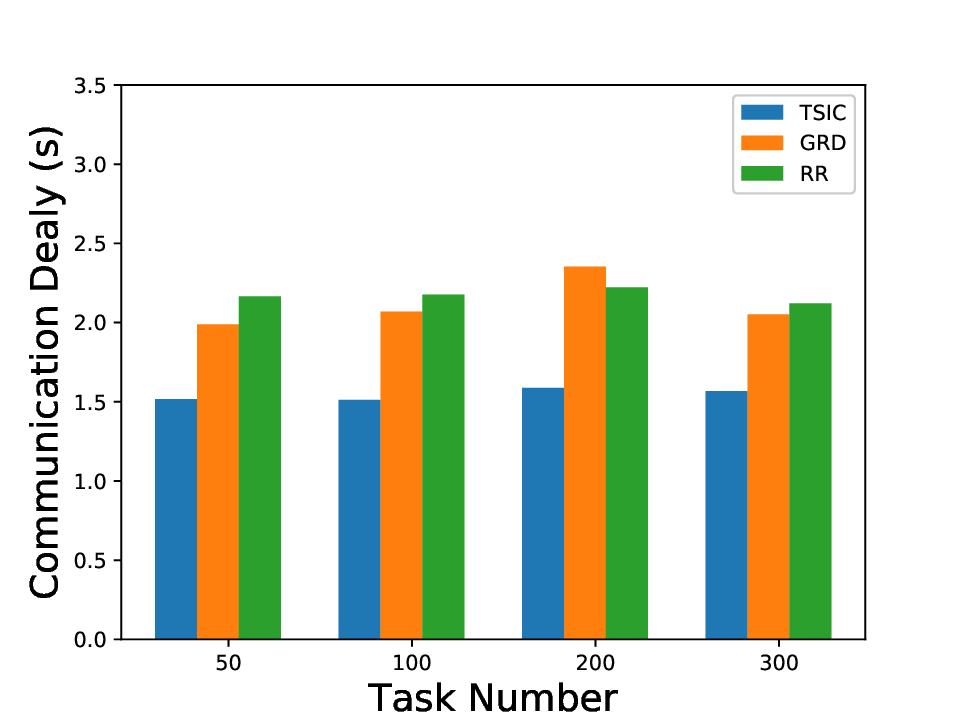}
    \label{fig:task_number_communication_delay}
  \end{minipage}
  }
	\hspace{\fill}
	\subfigure[Waiting Delay]{
	\begin{minipage}[t]{0.22\textwidth}
		\centering
		\includegraphics[width=1\textwidth]{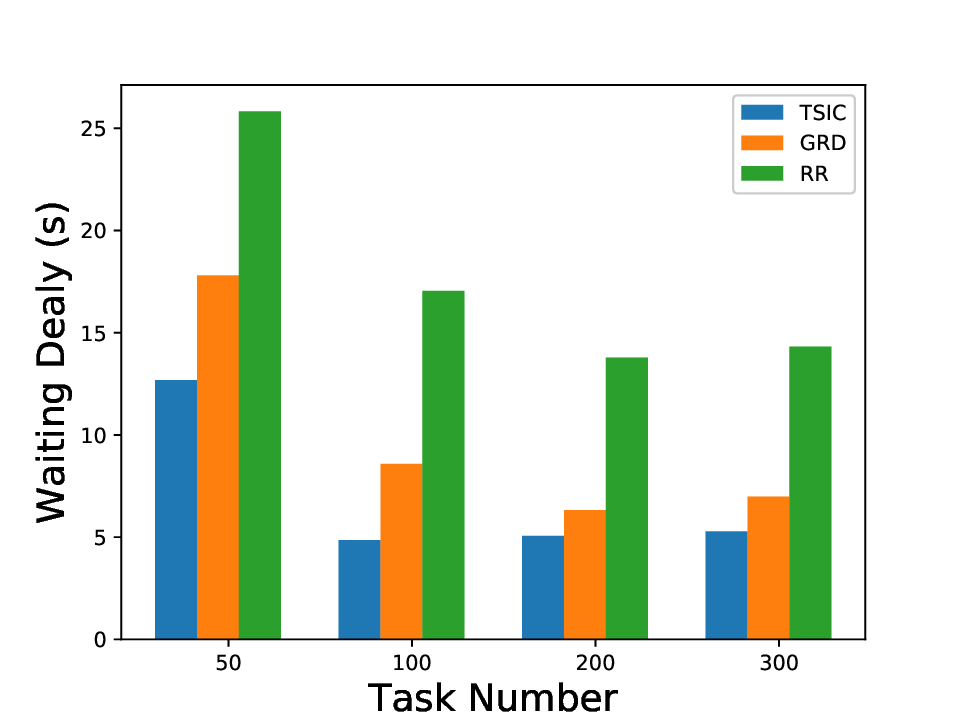}
		\label{fig:task_number_waiting_delay}
	\end{minipage}
        }
	\hspace{\fill}
	\subfigure[Computation Delay]{
  \begin{minipage}[t]{0.22\textwidth}
    \centering
    \includegraphics[width=1\textwidth]{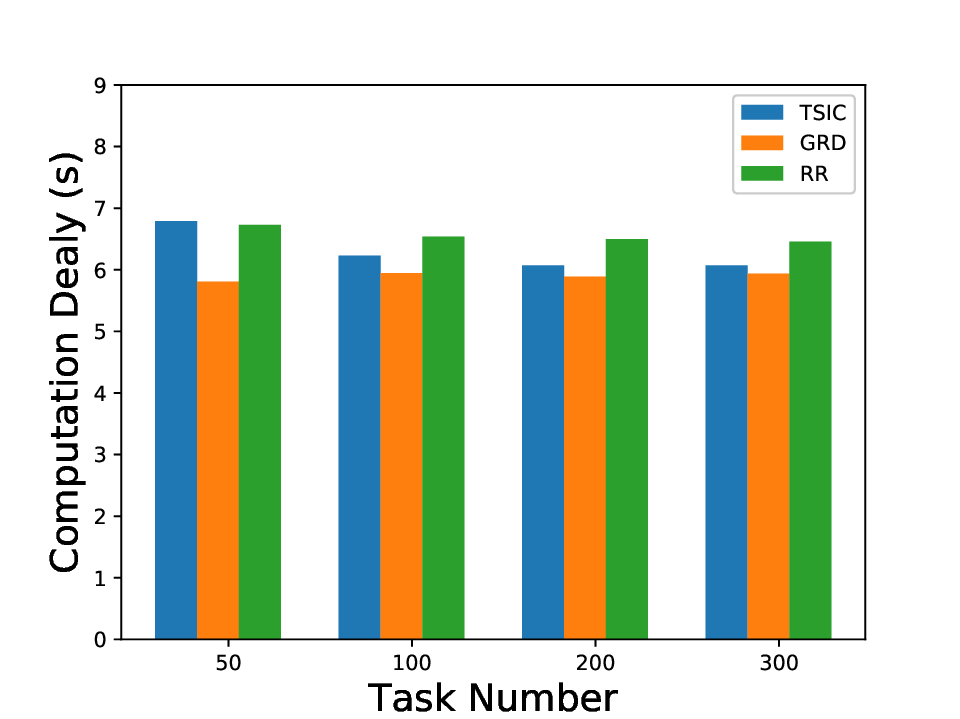}
    \label{fig:task_number_computation_delay}
  \end{minipage}
  }
   \hspace{\fill}
	\subfigure[Total Delay]{
  \begin{minipage}[t]{0.22\textwidth}
    \centering
    \includegraphics[width=1\textwidth]{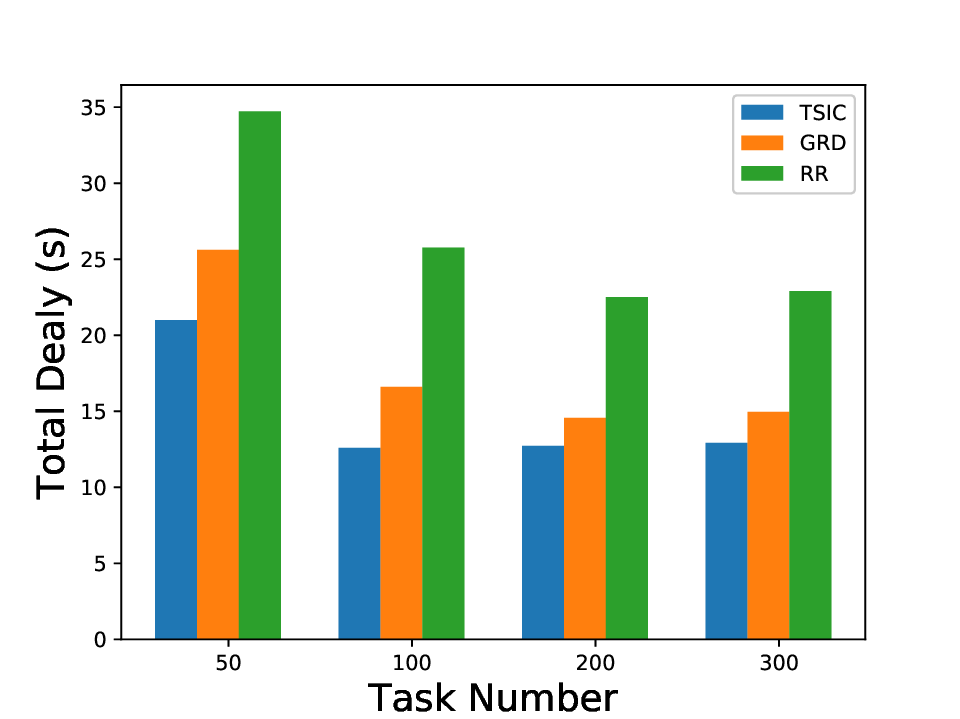}
    \label{fig:task_number_total_delay}
  \end{minipage}
  }
	\caption{Performance with Different Task Number}
	\label{fig:task_number}
\end{figure*}

\subsection{Experimental Settings}
\label{subsection_parameter}

The Raspberry Pis are evenly deployed in our laboratory according to the rectangle shape. All the Raspberry Pis and PCs are connected to a wireless router via WiFi. Since the distance between the Raspberry Pi and the wireless router is different, the transmission quality and speed of each Raspberry Pi are different. The different delays are all recorded directly in the experiment.

The images used in the experiments are built based on different Python OpenCV Docker images \cite{sgtwilko,mohaseeb}. Some Python code is written based on these images to perform picture-processing tasks like graying and compression. The sizes of the built images range from 253.07 MB to 458.73 MB. Each node has several random images located on it when initialized. Moreover, the $\epsilon$ in action selection is set to 0.5. The $\gamma$ is set to 0.5. The caching update frequency is set to 10, and the target network update frequency is set to 5.

\subsection{Experimental Results}
\label{subsection_experimental}


\textbf{Different LFU Memory Size}: Fig. \ref{fig:lfu_size} shows the communication delay, waiting delay, computation delay, and total delay with different LFU memory sizes. It can be concluded that the Adaptive LFU (ADP) performance is better than the fixed-size LFU. There is not much difference among the different LFU sizes for communication delay. The reason is that the proportion of communication delay is relatively small and will not be the focus of the learning process. 

For the waiting delay, it will be more significant if there are more images to be pulled. As the LFU memory size increases, the number of images that can be stored on each node grows. Then the number of images that need to be pulled is less, and the delay is reduced. To better reflect the impact of LFU size, the storage capacity of each node is limited artificially. With the limitation, the node with the most minor available storage can only accommodate up to 10 miniature images. As a result, the fixed-size LFU algorithm can easily reach the bottleneck and cannot fully utilize the resources of each node. Moreover, the image size-weighted LFU also performs better than the LFU algorithm only based on frequency (ADP-FRQ in the figure) because those images with a larger size but a lower frequency will not be removed frequently.

Finally, our algorithm will be more inclined to select nodes with more computation resources for computation delay. Generally, nodes with more available storage resources have more available computation resources. Overall, from the overall experiments, the total delay of the adaptive LFU is minimal.

\textbf{Different Node Number}: The performance of different node number is illustrated in Fig. \ref{fig:node_number}. GRD means the greedy algorithm, and RR means the round-robin algorithm. Fig. \ref{fig:node_number_communication_delay} demonstrates that the communication delay of TSIC is less than GRD and RR. Besides, as the number of nodes increases, the communication delay does not change much since it is unrelated to the number of nodes.

In Fig. \ref{fig:node_number_waiting_delay}, the waiting delay of these algorithms is ordered as TSIC $<$ GRD $<$ RR. TSIC makes caching decisions more effectively, and the caching time is effectively reduced, which is an essential part of the waiting time. Besides, the total storage capacity increase as the number of nodes increases. Then the images will not be removed frequently with the same number of tasks. As a result, the waiting delay gradually decreases as the number of nodes increases.

The computation delay of different algorithms is shown in Fig. \ref{fig:node_number_computation_delay}. In most cases, the greedy algorithm has the least computation delay. The reason is that the greedy algorithm always selects the node with the most available computation resources. However, the difference between the computation delay of TSIC and GRD is tiny. So, as a result, in Fig. \ref{fig:node_number_total_delay}, the total delay of TSIC is the smallest.

\textbf{Different Task Number}: The performance with different task number is shown in Fig. \ref{fig:task_number}. As shown in Fig. \ref{fig:task_number_communication_delay}, the computation delay does not change much as the number of tasks increases since it is not affected by the number of tasks. The performance of the waiting delay is shown in Fig. \ref{fig:task_number_waiting_delay}. As the number of tasks increases, some of the most popular images have been cached on different nodes, and the image distribution will not change much. So when the number of tasks increases, the waiting delay decreases first and then stabilizes. Besides, the waiting delay of TSIC is the least.

As shown in Fig. \ref{fig:task_number_computation_delay}, the performance of computation delay is GRD $<$ TSIC $<$ RR. This is because the value of the computation delay is much smaller than the waiting delay, and the waiting delay is prioritized during training. Besides, the greedy algorithm prioritizes computation resources. Such a computation delay gap is acceptable because more waiting delay is saved. Moreover, as shown in Fig. \ref{fig:task_number_total_delay}, the total delay is TSIC $<$ GRD $<$ RR.

To sum up, the TSIC algorithm outperforms the GRD and RR algorithms by 15\% and 31\% on average regarding the total delay, respectively. Besides, the TSIC algorithm is better for waiting delay than the GRD and RR algorithms, 28\% and 43\% on average, respectively. In short, the TSIC algorithm outperforms the existing baseline approaches 23\% and 35\% on average in terms of total delay and waiting delay, respectively.

\section{Conclusion}
\label{section_conclusion}

This paper has modeled the joint task scheduling and image caching problem in EC as an MDP problem. First, the system model is defined, whose cost function consists of communication, waiting, and computation delays. Then, a deep Q-learning-based joint algorithm is proposed. A state-sharing multi-action Q-network is proposed to achieve better decision-making, and an adaptive LFU-based caching update algorithm is combined. Experiments with the real container system have shown that our algorithms substantially reduce the total delay and waiting delay compared with the baselines. Future work will consider the mobility of mobile users and the trade-off between edge nodes and remote clouds.

\bibliographystyle{IEEEtran}
\bibliography{joint_scheduling}

\end{document}